\documentclass{LMCS}
\usepackage{hyperref,enumerate,proof}

\def\eqalign#1{\null\,\vcenter{\openup\jot
 \ialign{\strut\hfil$\displaystyle{##}$&$\displaystyle{{}##}$\hfil
     \crcr#1\crcr}}\,}
\def\rbox#1{\mbox{\rm #1}}
\def\:{\mathbin{\,:\,}}
\def\.{\mathbin{\,.\,}}
\def\RIghtarrow{\enspace\Rightarrow\enspace}
\def\aNd{\enspace\&\enspace}

\def\ebox#1{\enspace\hbox{\!}\enspace}

\newcommand{\Na}{{I\kern-4pt N}}

\def\doi{4 (1:1) 2008}
\lmcsheading%
{\doi}
{1--19}
{}
{}
{Nov.~21, 2006}
{Jan.~\phantom{0}7, 2008}
{}   

\begin{document}

\title[Hilbert-style Pure Type Systems?]{Are there Hilbert-style Pure 
  Type Systems?}

\author[M.~W.~Bunder]{Martin W.~Bunder\rsuper a}	
\address{{\lsuper a}School of Mathematics and Applied Statistics, 
University of Wollongong, Wollongong, NSW 2522, Australia}	 
\email{martin\_bunder@uow.edu.au}  
\thanks{{\lsuper{a,b}}The authors would like to thank the anonymous 
referees for their useful comments.}	

\author[W.~.J.~M.~Dekkers]{Wil J.~M.~Dekkers\rsuper b}	
\address{{\lsuper b}Department of Computer Science, Radboud 
University Nijmegen, Toernooiveld 1, 6525 ED Nij\-megen, The Netherlands}
\email{wil@cs.ru.nl}

\keywords{Hilbert-style logics, pure type systems, type theory, lambda 
calculus, illative combinatory logic}
\subjclass{F.4.1}
\amsclass{03B40, 03B70, 68N18}


\begin{abstract}
  For many a natural deduction style logic there is a Hilbert-style
  logic that is equivalent to it in that it has the same theorems
  (i.e.\ valid judgements $\Gamma\!\vdash\!P$ where
  $\Gamma=\emptyset)$. For intuitionistic implicational logic, the
  axioms of the equivalent Hilbert-style logic can be propositions
  which are also known as the types of the combinators ${\bf I}$,
  ${\bf K}$ and ${\bf S}$.

  Natural deduction versions of illative combinatory logics have
  formulations with axioms that are actual type statements for ${\bf
    I}$, ${\bf K}$ and ${\bf S}$. As pure type systems (PTSs) are, in
  a sense, equivalent to systems of illative combinatory logic, it
  might be thought that Hilbert style PTSs (HPTSs) could be based in a
  similar way.

  This paper shows that some PTSs have very trivial equivalent HPTSs,
  with only the axioms as theorems and that for many PTSs no
  equivalent HPTSs can exist.  Most commonly used PTSs belong to these
  two classes.

  For some PTSs however, including $\lambda^ *$ and the PTS at the
  basis of the proof assistant Coq, there is a nontrivial equivalent
  HPTS, with axioms that are type statements for ${\bf I, K}$ and\
  ${\bf S}$.
\end{abstract}

\maketitle

\section*{Introduction}\label{S:one}

  \noindent Most early logical systems (for propositional and
  predicate logic) allowed no hypotheses and so had no rules for
  introducing or cancelling them.  These could be represented by a
  finite set of axiom schemes and rules of inference such as modus
  ponens and generalisation.

  Later natural deduction systems which did allow hypotheses had fewer
  axiom schemes but required introduction and elimination rules for
  hypotheses. Herbrand showed that classical Hilbert style and natural
  deduction style propositional and predicate logics had the same
  theorems (i.e. judgements with empty contexts).

  Pure type systems (PTSs), defined below, have two rules that
  introduce hypotheses and two that cancel them. In this paper we
  answer a question of Fairouz Kamareddine ``Are there Hilbert style
  PTSs?". When we define Hilbert style PTSs (HPTSs) as PTSs with empty
  contexts, with a finite set of extra axiom schemes, with $s_1, s_2,
  s_3,\dots $ representing arbitrary sorts, and some extra rules, it is
  obvious that there are HPTSs. We will be interested in whether, for
  PTSs, there are theorem equivalent HPTSs. We will answer this
  question for a number of classes of PTSs which include all the PTSs,
  from the standard literature, that we have examined. The methods we
  use, for proving that a HPTS is equivalent to a PTS, are along the
  lines of those of Herbrand, but rather more complex.

  Just as combinator based programming languages, requiring no free,
  or in fact, no variables, have proved useful in practice, perhaps an
  HPTS, which also requires no (free) variables, that is theorem
  equivalent to a PTS may be useful. Also, perhaps some
  metatheoretical results may be proved more easily for an HPTS than
  for the equivalent PTS.

\section{Pure Type Systems}\label{S:PTS}

  \noindent Each Pure Type System (PTS) $\lambda X$ has a set of
  variables $V$, a set of constants ${\mathcal C}$, a set of ``sorts"
  $\mathcal S\subseteq{\mathcal C}$. It has a class of pseudoterms
  given by ${\mathcal T}\!=\!V\,|\,{\mathcal C}\,|\,(\Pi
  V\!:\!\mathcal{T.T})\,|\,(\lambda V\!:\!\mathcal{T.T})\,|\,\mathcal{TT}$. If
  $M$ and $N$ are pseudoterms, $M\:A$ is a statement, $\Gamma$ is a
  context if it is a sequence of statements; $\Gamma\vdash M\:A$ is
  then called a judgement. A PTS has a set of axioms ${\mathcal A}$
  each of the form $c:s$ where $c\in{\mathcal C}$ and $s\in\mathcal
  S$. Then it has a set $\mathcal R$ of triples
  $(s_1,s_2,s_3)\in\mathcal {S}$$^3$, which determine under what
  conditions a term $\Pi x{:}A.B$ is in a sort.  Most PTSs are known
  by a ``specification" $(\mathcal S, {\mathcal A},\mathcal R)$ (as
  usually ${\mathcal C}=\mathcal S$).

  The PTS postulates are as follows:
\smallskip
{\small\[\begin{tabular}{lllllll}
(axiom)&&& $\displaystyle{{c:s\in {\mathcal A}\over\vdash c:s}}$\cr\cr
(start)&&&$\displaystyle{\Gamma\vdash A\:s\over\Gamma,x\:A\vdash x\:A}$\cr\cr
(weakening)&&& $\displaystyle
{{\Gamma\vdash M\:B\qquad\Gamma\vdash A\:s\qquad x\notin FV
(\Gamma)\over\Gamma, x\:A\vdash M\:B }}$\cr\cr
(application)&&& $\displaystyle{{\Gamma\vdash M\:(\Pi x{:}A.B)\qquad
\Gamma\vdash N\:A\over\Gamma\vdash MN:B[x:=N]}}$\cr\cr
(abstraction)&&&$\displaystyle{{\Gamma,x\:A\vdash
M\:B\qquad\Gamma\vdash(\Pi x{:}A.B):s\over\Gamma\vdash(\lambda
x{:}A.M)\:(\Pi x{:}A.B)}}$\cr\cr
(product)&&&$\displaystyle{{\Gamma,x\:A\vdash B\:s_2
\qquad\Gamma\vdash A\:s_1\qquad (s_1,s_2,s_3)\in {\mathcal R}\over
\Gamma\vdash(\Pi x{:}A.B):s_3}}$\cr\cr
(conversion)&&&$\displaystyle{{\Gamma\vdash M\:A\qquad\Gamma\vdash
B\:s\qquad A=_{\beta} B\over\Gamma\vdash M\:B}}.$\cr
\end{tabular}
\]}\smallskip

  \noindent When there are two judgements as premises in a rule, we call the
  left one the \emph{major premise} and the right one the \emph{minor
    premise}.

  Later we will need the following definition:

\begin{defi}[{Inhabited and Normal Form Inhabited Sorts}]\hfill

  $s$ is an \emph{inhabited sort} $(s\in\mathcal I)$ if $\vdash A\:s$
  for some $A$.

  $s$ is a \emph{normal form inhabited sort} $(s\in\mathcal N)$ if
  for some term $A$ in normal form, $\vdash A\:s.$
\end{defi}

  The translation $[\ \ ]$ of Bunder and Dekkers [3]
  translates the pseudoterms and statements of PTSs into terms
  of illative combinatory logic (ICL) as follows:
\[[x] = x,\quad[c]=c,\quad [XY]=[X][Y]\]
\[[X\:A] = [A][X],\quad [\Pi x{:}X.Yx] ={\bf G} [X][Y]\quad (x\notin FV(XY))\]
  where ${\bf G}=\lambda xyz.\Xi x({\bf S}yz)$ (${\bf S}$ is the
  combinator equivalent to $\lambda xyz.xz(yz))$. Terms in ICL can be
  represented without any free variables at all using the combinators
  ${\bf S}$ and ${\bf K}$ (equivalent to $\lambda xy.x$). $\Xi xy$
  represents roughly $(\forall u\in x) y(u)$ or $x\subseteq y$.

  ICL, designed as a foundation for logic and mathematics, has a rule
  like (abstraction) which was derived in Bunder [2] from a set of
  axioms. In Section 6 we will see how the methods developed there
  lead to the ones used here. The main difference between PTSs and
  standard ICLs, other than the lack of distinction between terms and
  types, lies in the (abstraction) rule. The direct counterpart to the
  ICL rule would have $\Gamma\vdash A\:s$, for $\Gamma\vdash(\Pi
  x{:}A.B):s$.  This is the most important factor in making it
  difficult to have equivalent Hilbert-style PTSs.

\section{Hilbert-style PTSs}\label{S:HSPTS}

 \noindent We define Hilbert-style PTSs as follows:

\begin{defi}[HPTS]

  Each \emph{Hilbert style Pure Type System} (HPTS) has $V$, $\mathcal
  C$, $\mathcal S$, $\mathcal T$, statements, contexts and judgements
  as for PTSs, except that the contexts are always empty. A HPTS has a
  set of sorts $\mathcal S$ and a set of axioms ${\mathcal A}$, as for
  PTSs, and an additional finite set ${\mathcal B}$ of axiom schemes
  in which ``sort variables" can be replaced by sorts.  Most HPTSs are
  known by a ``specification" $(\mathcal S, {\mathcal A},\mathcal B)$
  (as usually ${\mathcal C}=\mathcal S$). A HPTS has the PTS
  (application) and (conversion) rules (with empty contexts) as well
  as:
\bigskip
{\small\[\begin{tabular}{lllllll}
(type reduction)&&& $\displaystyle{{\vdash M\:A\qquad
   A\rightarrow_{\beta} B\over\vdash M\:B.}}$\cr\cr
(subject reduction)&&&$\displaystyle{{\vdash M\:A\qquad
   M\rightarrow_{\beta} N\over\vdash N\:A.}}$\cr\cr
\end{tabular}
\]}
\end{defi}
  Note the latter rules are derivable for all PTSs, for HPTSs neither
  is, even using (conversion).

\begin{defi}[Equivalent  HPTS]

  If $\lambda X$ is a PTS with specification $(\mathcal S,{\mathcal
  A,}\mathcal R)$, a HPTS $\lambda X^h$, with specification
  $(\mathcal S,{\mathcal A},\mathcal B)$ will be \emph{equivalent} if
\[(\forall M,A)(\vdash^X M\:A\enspace\Leftrightarrow\enspace\vdash^{X^h} M\:A).\]
  Here $\vdash^X$ stands for provability in $\lambda X$ and $\vdash
  ^{X^h}$ in $\lambda X^h$.  If the PTS is arbitrary or obvious from
  the context we use $\vdash$ and $\vdash^h$. $\mathcal B$ will a
  function of $\mathcal R$, i.e. it will include axioms such as
  $\vdash^h[\lambda u{:}s_1 .\lambda v{:}(\Pi x{:}u . s_2) .\Pi
  x{:}u.vx]\:[\Pi u{:}s_1.\Pi v{:}(\Pi x{:}u. s_2).s_3]$ if
  $(s_1,s_2,s_3)\in\mathcal R.$
\end{defi}

  Below are some PTSs that have been studied in the literature
  (particularly Barendregt [1] and Geuvers [4]).

  In $\lambda^\tau ,\mathcal S=\{ *\}$, ${\mathcal C}=\{*,0\}$, in
  all other cases ${\mathcal C} =\mathcal S$ consists of all the
  constants visible in ${\mathcal A}$ and $\mathcal R$.  $(s_1, s_2)$
  is used as an abbreviation for $(s_1, s_2, s_2)$.
\bigskip
{\small\[\begin{tabular}{llllllllll}
$\lambda^{\rightarrow}$ & ${\mathcal A}=\{*:\square\}$&
$\mathcal R=\{(*,*)\}$\cr\cr
$\lambda^\tau$ & ${\mathcal A}=\{0:*\}$&
$\mathcal R=\{(*,*)\}$\cr\cr
$\lambda ^*$ & ${\mathcal A}=\{*:*\}$&
$\mathcal R=\{(*,*)\}$\cr\cr
$\lambda 2$ & ${\mathcal A}=\{*:\square\}$&
$\mathcal R=\{(*,*), (\square,*)\}$\cr\cr
$\lambda $P & ${\mathcal A}=\{*:\square\}$&
$\mathcal R =\{(*,*),(*,\square)\}$\cr\cr
$\lambda\underline{\omega}$ & ${\mathcal A}=\{*:\square\}$&
$\mathcal R=\{(*,*), (\square,\square)\}$\cr\cr
$\lambda\omega$ & ${\mathcal A}=\{*:\square\}$&
$\mathcal R=\{(*,*),(\square, *), (\square,\square)\}$\cr\cr
$\lambda $P2 & ${\mathcal A}=\{*:\square\}$&
$\mathcal R=\{(*,*),(\square, *), (*,\square)\}$\cr\cr
$\lambda $P$\underline{\omega}$ & ${\mathcal A}=\{*:\square\}$&
$\mathcal R=\{(*,*),(*,\square), (\square,\square)\}$\cr\cr
$\lambda $P$\omega=\lambda $C & ${\mathcal A}=\{*:\square\}$&
$\mathcal R=\{(*,*),(*,\square),(\square,*),(\square,\square)\}$\cr\cr
$\lambda$AUT-68 & ${\mathcal A}=\{*:\square\}$&
$\mathcal R=\{(*,*),(*,\square,\triangle),
(\square,*,\triangle),(\square,\square,\triangle),(*,\triangle),
(\square,\triangle)\} $\cr\cr
$\lambda$AUT-QE & ${\mathcal A}=\{*:\square\}$& $\mathcal
R=\{(*,*),(*,\square),
(\square,*,\triangle),(\square,\square,\triangle),(*,\triangle),
(\square,\triangle)\}$\cr\cr
$\lambda$PAL & ${\mathcal A}=\{*:\square\}$& $\mathcal
R=\{(*,*,\triangle),(*,\square,\triangle),
(\square,*,\triangle),(\square,\square,\triangle),(*,\triangle),
(\square,\triangle)\}$\cr\cr
$\lambda $U & ${\mathcal
A}=\{*:\square,\square:\triangle\}$& $\mathcal R=\{(*,*),(\square,*),
(\square,\square),(\triangle,\square),(\triangle,*)\}$\cr\cr
$\lambda $HOL & ${\mathcal A}=\{*:\square,\square:\triangle\}$& $\mathcal R=\{(*,*),(\square,*),(\square,\square)\}$\cr\cr
\end{tabular}
\]}
  The PTS used in the proof assistant Coq we will call $\lambda\rbox{Coq}$.
  It has as axioms:
\[\vdash *_p\:\square_1\ebox{,}\vdash *_s\:\square_1\ebox{,}\forall
  i\in\Na\vdash\square_i\:\square_{i+1}.
\]
  More axioms are generated by
\[A\:B, B\:C\in {\mathcal A}\RIghtarrow A\:C\in{\mathcal A}.\]
  In early versions ${\mathcal R}$ is given by
\[(*_s,*_s),(*_p,*_p),(*_s,*_p),(*_p,*_s),(*_p,\square_i),(*_s,\square_i),(\square_i,*_p),(\square_i,*_s),(\square_i,
\square_j,\square_{max(i,j)})\in\mathcal R.\] for all $i,j\in\Na$.
  Coq 8.0 replaces $(\square_i,*_s)\in {\mathcal R}$ by $(\square_i,
  *_s,\square_i)\in {\mathcal R}$.

  We will be able to determine whether or not there are equivalent
  HPTSs for all of the above.

\section{Some PTS Lemmas and Definitions}\label{S:PTSLD}

 \noindent We now state a number of standard lemmas for PTSs.  Most
  proofs can be found in Barendregt [1] or Bunder and Dekkers [3].

\begin{lem}[Free Variable Lemma]\label{L:one}

  If $x_1\:A_1,\ldots,x_n\:A_n\vdash M\:B$, then
\begin{enumerate}[\em(i)]
\item $x_1,\ldots,x_n$ are distinct;
\item $FV(M,B)\subseteq\{x_1,\ldots,x_n\}$;
\item $FV(A_i)\subseteq\{x_1\ldots,x_{i-1}\}$ for $1\leq i\leq n$.\qed
\end{enumerate}
\end{lem}

\begin{lem}[Substitution Lemma]\label{L:two}

  If $\Gamma_1,x\:A$, $\Gamma_2\vdash M\:B$ and $\Gamma_1\vdash N\:A$
  then $\Gamma_1,\Gamma_2[x:=N]\vdash M[x:=N]\:B[x:=N]$.\qed
\end{lem}

\begin{lem}[Condensing Lemma]\label{L:three}

  If $\Gamma_1, x\:A,\Gamma_2\vdash M\:B$, where $x\notin FV(\Gamma_2,
  M,B)$, then $\Gamma_1,\Gamma_2\vdash M\:B$.\qed
\end{lem}

\begin{lem}[Generation Lemma]\label{L:four}

  Let $\Gamma\vdash M\:B$.  Then
\begin{enumerate}[\em(i)]
\item
$M\equiv c\in\mathcal C\RIghtarrow (\exists s\in
{\mathcal S})\ B =_{\beta} s\aNd c:s\in {\mathcal A}$;

\item
$M\equiv x\RIghtarrow (\exists C)\ B =_{\beta} C\aNd x\:C\in
\Gamma$;

\item
$M\equiv\Pi x{:}C.D\RIghtarrow (\exists s_1,s_2,s_3)\
(s_1,s_2,s_3)\in{\mathcal R}\aNd\Gamma\vdash C:s_1$\hfill\break
$\phantom{M\equiv\Pi x{:}C.D\RIghtarrow (\exists s_1,s_2,s_3)\
(s_1,s_2,s_3)\in{\mathcal R}}
\aNd\Gamma, x\:C\vdash D:s_2\aNd B=_{\beta} s_3$;

\item
$M\equiv\lambda x{:}C.N\RIghtarrow (\exists s\in{\mathcal S})\,
 (\exists D)\Gamma,x\:C\vdash N:D\aNd B=_{\beta}\Pi
 x{:}C.D$\hfill\break
$\phantom{M\equiv\lambda x{:}C.N\RIghtarrow (\exists s\in{\mathcal S})\,
 (\exists D)\Gamma,x\:C\vdash N:D}
\aNd\Gamma\vdash(\Pi x{:}C.D):s$;

\item
$M\equiv PQ\RIghtarrow (\exists C,D)\Gamma\vdash P:\Pi
x{:}C.D\aNd\Gamma\vdash Q:C\aNd B=_{\beta} D[x:=Q]$.
\end{enumerate}
  In each case the derivations, of the judgements of the form
  $\Gamma\vdash R:E$ in \emph{(iii)} to \emph{(v)}, are shorter than
  that of $\Gamma\vdash M\:B$.\qed
\end{lem}

\begin{lem}[Correctness of Types Lemma]\label{L:five}

  If $\Gamma\vdash M\:B$ then $(\exists s\in\mathcal S)$ $[B\equiv
  s$ or $\Gamma\vdash B\:s]$.\qed
\end{lem}

\begin{lem}[Subject and Type Reduction Lemma]\label{L:six}

  If $\Gamma\vdash M\:B$, then
\begin{enumerate}[\em(i)]
\item $M\rightarrow\hspace{-3mm}\rightarrow_{\beta} N$ implies
  $\Gamma\vdash N:B$,
\item and $B\rightarrow\hspace{-3mm}\rightarrow_{\beta}
 A$ implies $\Gamma\vdash M\:A$.\qed
\end{enumerate}
\end{lem}

\begin{lem}[Start Lemma]\label{L:seven}

  If $\Gamma\vdash M\:B$, then
\begin{enumerate}[\em(i)]
\item $(c:s)\in\mathcal A$ implies $\Gamma\vdash c\:s$,
\item $\Gamma\equiv x_1\:A_1,\dots ,x_n:A_n$implies that for $0\le i
  <n$ there is an $s\in\mathcal S$ such that $x_1\:A_1,\dots ,x_i\:A_i
 \vdash A_{i+1}\:s$.\qed
\end{enumerate}
\end{lem}

\section{PTSs where \texorpdfstring{$\mathcal A$}{A} is the Set of 
  Theorems}\label{S:PTSST}

 \noindent The following lemma specifies a set of PTSs whose axioms
 are its only theorems.  The equivalent HPTS is then trivially one
 with no extra axioms, i.e. with ${\mathcal B}=\emptyset$.

\begin{lem}\label{L:eight} In a PTS satisfying
\[(\forall c,s_1)\bigl((c:s_1)\in {\mathcal A}\RIghtarrow\sim(\exists
  s_2,s_3)[(s_1,s_2,s_3)\in\mathcal R]\bigr)\eqno{(\$)}
\]
  we have $\vdash M\:A\enspace\Leftrightarrow\enspace M\:A\in{\mathcal A}$.
\end{lem}

\proof We show $M\:A\in\mathcal A$ by induction on the derivation of
\begin{equation}\label{EQ:ma}
 \vdash M\:A.
\end{equation}

  (\ref{EQ:ma}) clearly does not come by (start) or (weakening).

  If (\ref{EQ:ma}) comes by (application) from
\[\vdash P:(\Pi x{:}C. D)\quad\mbox{and}\quad\vdash Q:C\]
  where $M\equiv PQ$ and $A\equiv D[x:=Q]$, we have by the induction
  hypothesis $P\:(\Pi x{:}C. D)\in {\mathcal A}$, which is impossible.

  If (\ref{EQ:ma}) comes by (abstraction) from
\[x\:B\vdash N:C\quad\mbox{and}\quad\vdash(\Pi x{:}B. C):s\] 
  where $M\equiv\lambda x{:}B.N$ and $A\equiv\Pi x{:}B.C$, then by
  the induction hypothesis $(\Pi x{:}B.  C):s\in {\mathcal A}$, which
  is impossible.

  If (\ref{EQ:ma}) comes by (product) from
\[x\:B\vdash C:s_2\quad\mbox{,}\quad\vdash B\:s_1
                 \quad\mbox{and}\quad(s_1,s_2,s_3)\in\mathcal R
\]
  where $M\equiv\Pi x{:}B.C$ and $A\equiv s_3$, then, by the
  induction hypothesis, $(B\:s_1)\in {\mathcal A}$, which is impossible
  by (\$).

  If (\ref{EQ:ma}) comes by (conversion) from
\[\vdash M\:B\quad\mbox{,}\quad\vdash A\:s
                 \quad\mbox{and}\quad A=_\beta B
\]
  then by the induction hypothesis $(M\:B), (A\:s)\in{\mathcal A}$.
  However then $A$ and $B$ must be in normal form and so $A\equiv B$
  and $(M\:A)\in{\mathcal A}$.

  If (\ref{EQ:ma}) is an axiom, the result holds trivially.\qed

  This implies the following theorem and corollary.

\begin{thm}\label{T:nine} 
  A PTS satisfying  \emph{(\$)} has an equivalent HPTS, with
  $\mathcal B =\emptyset,$ but this is trivial in that it has only
  its axioms as theorems.\qed
\end{thm}

\begin{cor}\label{C:ten}
  $\lambda^{\rightarrow}$ and $\lambda $P each have an equivalent
  HPTS, but $\vdash\ast\:\square$ is the only theorem of both
  systems.\qed
\end{cor}

\section{PTSs with no Equivalent HPTS}\label{S:PTSnoHPTS}

 \noindent In $\lambda^{\rightarrow}$ and $\lambda$P there is no term
  $A$ such that $\vdash A:\ast$ and the only theorem is $\vdash\ast :
 \square$.

  We can show, by a single (product) rule preceeded by two uses of an
  axiom and a (start) or (weakening) rule, that in the other PTSs,
  given in Section 2, there are theorems that are not axioms. Most of
  these are given below.

\begin{lem}\label{L:eleven}\hfill

\begin{enumerate}[\em(i)]
\item In $\lambda^\tau$  we have $\vdash(\Pi x{:}0.0):\ast$.
\item In $\lambda^\ast$, $\lambda2$, $\lambda\rbox{P2}$, $\lambda\omega$,
  $\lambda\rbox{C}$, $\lambda\rbox{U}$ and $\lambda\rbox{HOL}$, we have
  $\vdash(\Pi x{:}\ast.x):\ast$.
\item In $\lambda\underline{\omega}$ and
  $\lambda\rbox{P}\underline{\omega}$, we have
  $\vdash (\Pi x{:}\ast.\ast):\square$.
\item In $\lambda\rbox{AUT-68}$, $\lambda\rbox{AUT-QE}$ and
  $\lambda\rbox{PAL}$ we have
  $\vdash (\Pi x{:}\ast.\ast):\triangle$.\qed
\end{enumerate}
\end{lem}

  We now give a condition under which, in a PTS, certain sorts have an
  infinite number of inhabitants of the form $\Pi x{:}A.B$ that are
  not substitution instances of each other.  We show later that many
  PTSs with this property cannot be equivalent to HPTSs.

\begin{lem}\label{L:twelve} Assume that in a PTS there is a finite
  sequence $s_1, s_2,\dots ,s_n\in\mathcal S$ such that:
\[(\exists n\in\Na)\ n>1\aNd s_1=s_n\in {\mathcal N}\aNd 
  (\forall i)\bigl(1\leq i < n\Rightarrow (\exists s'\in {\mathcal I}
\aNd(s', s_i, s_{i+1})\in {\mathcal R})\bigr)\eqno{(\$s_1,\dots,s_n)}
\]
  then
\[\vdash (\Pi x{:}A.B):s_1\]
  for an infinite number of $\beta$-distinct terms $\Pi x{:}A.B$
  which are not ($s_i$ for $s_j$) substitution instances of each
  other.
\end{lem}

\proof Assume that we have ($\$s_1,\dots,s_n$) for $s_1, s_2,\dots \in
 \mathcal S$.

  As $s_1\in {\mathcal N}$ we have, for some $A_1$, in normal form
\[\vdash A_1:s_1.\]
  Now we show, by induction on $i$ that, for $1 < i\leq n$, there is
  a $B_{i-1}$ and an $A_i =\Pi x_{i-1}{:}B_{i-1} . A_{i-1}$ such
  that
\begin{equation}\label{EQ:two}
 \vdash A_i:s_i.
\end{equation}
  For each $i$ we have, by ($\$s_1,\dots,s_n$), an $s'$ such that $(s',
  s_{i-1}, s_i)\in {\mathcal R}$ and a $B_{i-1}$ such that
\[\vdash B_{i-1}:s'.\]
  When $i=2$ we have $\vdash A_1:s_1$ above, otherwise we have $\vdash
  A_{i-1}:s_{i-1}$ by the induction hypothesis. By (weakening) we have
\[x_{i-1}:B_{i-1}\vdash A_{i-1}:s_{i-1}\]
  and by (product) we have (\ref{EQ:two}).  So (\ref{EQ:two}) holds for
  $1\leq i\leq n$ and, as we have $s_1=s_n$,
\[\vdash A_n:s_1.\]
  Repeating the above, with $A_n$ for $A_1$, we get $\vdash
  A_{2n-1}:s_1$ and similarly $\vdash A_{3n-2}:s_1$,\dots

  If $A_{in-i+1}=_\beta A_{jn-j+1}$ for $i<j$, then
\[\Pi x_{in-i}{:}B_{in-i}.A_{in-i}=_\beta
 \Pi x_{jn-j}{:}B_{jn-j}.A_{jn-j}
\]
  and so $A_{in-i}=_\beta A_{jn-j}$ and eventually
\[A_1 = _\beta A_{(j-i)(n-1)+1}.\]
  But $A_1$ is a proper part of $A_{(j-i)(n-1)+1}$ and is in normal
  form, which is impossible.  Hence $A_n, A_{2n-1},A_{3n-2},\ldots$
  are $\beta$-distinct inhabitants of $s$ all of the form $\Pi
  x{:}A.B$, which are not substitution instances of each other.\qed
\nobreak

  ($\$s_1,\dots,s_n$) is satisfied for many sequences $s_1, s_2,\dots .,s_n$
  and many PTSs.  Here we list one such sequence and sort for most of
  the PTSs given in Section 2.

\begin{lem}\label{L:thirteen}\hfill

\begin{enumerate}[\em(i)]
\item $\lambda^\tau$, $\lambda^\ast$ and $\lambda2$ satisfy
  \emph{($\$\ast,\ast$)}. 
\item $\lambda\underline{\omega}$, $\lambda\omega$,
  $\lambda\rbox{P2}$, $\lambda\rbox{P}\underline{\omega}$,
  $\lambda\rbox{C}$, $\lambda\rbox{U}$ and $\lambda\rbox{HOL}$
  satisfy \emph{($\$\square,\square$)}.
\item $\lambda\rbox{AUT-68}$, $\lambda\rbox{AUT-QE}$ and
  $\lambda\rbox{PAL}$ satisfy \emph{($\$\triangle,\triangle$)}.
\end{enumerate}
\end{lem}
\vfill\eject

\proof\hfill

\begin{enumerate}[(i)]
\item By Lemma \ref{L:eleven}(i), (ii) with $s' =\ast$.
\item For $\lambda\rbox{P2}$, by Lemma \ref{L:eleven}(ii), with $s'
  =*$.  For the others with $s' =\square$.
\item By Lemma \ref{L:eleven}(iv) with $s' =\square$.\qed
\end{enumerate}

  \noindent Now we can prove the main result in the section.

\begin{thm}\label{T:fourteen} 
  If, in a PTS $\lambda X$, \emph{($\$s_1,\dots ,s_n$)} holds for some
  $s_1,\dots ,s_n\in {\mathcal S}$ and
\[(\forall s_1', s_2', s_3')\bigl((s_1', s_2', s_3')\in {\mathcal R}
\Rightarrow (s_1:s_2')\not\in {\mathcal A}\bigr)\,,\eqno{(\$\$s_1)}
\]
  then there is no HPTS equivalent to $\lambda X$.
\end{thm}

\proof 
  By Lemma \ref{L:twelve}, if ($\$s_1,\dots ,s_n$) holds we have, for
  an infinite number of $\beta$ - distinct terms $\Pi x{:}A.B$, which
  are not substitution instances of each other
\[\vdash(\Pi x{:}A.B):s_1.\]

  Suppose that there is an equivalent $\lambda X^h$.

  As a HPTS has only a finite set of axioms $\mathcal B$, at least
  some must be derived, in $\lambda X^h$, by (application) and perhaps
  (conversion), (type reduction) and (subject reduction) from
\[\vdash^h P:(\Pi y{:}D.E)\quad\mbox{and}\quad\vdash^hQ:D\]
  where $PQ\rightarrow_{\beta}\Pi x{:}A.B$ and $E[y:=Q]
 \rightarrow_{\beta} s_1$.
  By the equivalence of $\lambda X$ and $\lambda X^h$ also:
\begin{equation}\label{EQ:three}
 \vdash P:(\Pi y{:}D.E)\quad\mbox{and}\quad\vdash Q:D.
\end{equation}
  So by correctness of types (Lemma \ref{L:five}), for some $s_3'\in
  {\mathcal S}$
\[\vdash(\Pi y{:}D.E):s_3'\]
  and by the Generation Lemma (Lemma \ref{L:four}(iii)) we have:
\[\vdash D\:s_1'\quad\mbox{and}\quad y\:D\vdash E\:s_2'\]
  where $(s_1',s_2',s_3')\in {\mathcal R}$.

  Now by the substitution lemma (Lemma \ref{L:two}), (\ref{EQ:three})
  and, if needed, subject reduction (Lemma \ref{L:six}) $\vdash
  s_1:s_2'$.

  By Lemma \ref{L:four}(i) this contradicts ($\$\$s_1$), so $\lambda X$ has no
  HPTS equivalent.\qed

\begin{thm}\label{T:fifteen}
  $\lambda^\tau$, $\lambda2$, $\lambda{\omega}$,
  $\lambda\underline{\omega}$, $\lambda\rbox{P2}$,
  $\lambda\rbox{P}\underline{\omega}$, $\lambda\rbox{C}$,
  $\lambda\rbox{AUT-68}$, $\lambda\rbox{AUT-QE}$, $\lambda\rbox{PAL}$,
  $\lambda\rbox{U}$ and $\lambda\rbox{HOL}$ have no equivalent HPTSs.
\end{thm}

\proof By Lemma \ref{L:thirteen} and Theorem \ref{T:fourteen}.\qed

  Note that ($\$\$s$) is not satisfied by any sort $s$ in
  $\lambda^{\ast}$ and $\lambda $Coq.  For $\lambda^{\rightarrow}$
  ($\$\$\ast$) holds but ($\$\ast,s_2,\dots ,\ast$) does not for any
  $s_2,\dots $ For $\lambda $P ($\$\$\ast$) fails, ($\$\$\square$) holds
  but ($\$\square,s_2,\dots ,\square$) does not for any $s_2,\dots $
\vfill\eject

\section{How to Prove (abstraction) and (product)}\label{S:AbsProd}

 \noindent In implicational logic the $\supset$-introduction rule is
  $\Gamma, A\vdash B\RIghtarrow\Gamma\vdash A\supset B$. The
  hypothesis A in $\Gamma, A\vdash B$ is cancelled in $\Gamma\vdash
  A\supset B$. This rule is proved in a Hilbert-style system by
  induction on the number of steps in a derivation that allows
  hypotheses. We assume that an hypothesis can be cancelled in the
  previous step (or steps) and use this to show it can be cancelled in
  the next.  In intuitionistic and classical implicational logic three
  cases are needed and each requires the Hilbert-style system to have
  a particular axiom or theorem.

  If the hypothesis $p$ is itself the step in the deduction we need
\[\vdash p\supset p.\]
  If the deduction step is an axiom or another hypothesis than $p$ we need
\[\vdash q\supset p\supset q.\]
  If the deduction step comes  by modus ponens from
\[p\vdash q\quad\mbox{and}\quad p\vdash q\supset r\ ,\]
  we need 
\[\vdash (p\supset q\supset r)\,\supset\, (p\supset q)\supset p\supset r.\]

  Note that the three theorems we require represent the simple types
  of the combinators ${\bf I}$, ${\bf K}$ and ${\bf S}$ (when
  $\supset$ is replaced by $\rightarrow$).

  In illative combinatory logic, the introduction rule for $\Xi$
  (restricted generality) is $\Gamma, Ax\vdash Bx\RIghtarrow\Gamma,
  {\bf L}A\vdash\Xi AB$, where ${\bf L}$ is a constant, $x\notin
  FV(\Gamma, A,B)$ and $Ax$ is the hypothesis being cancelled. In the
  proof of this rule in a Hilbert-style system,(see Bunder[2]), the
  first two cases are similar to those for the proof of implicational
  introduction. The third is the case where $\Gamma, Ax\vdash DM$ is
  derived from $\Gamma, Ax\vdash\Xi CD$ and $\Gamma, Ax\vdash CM$.
  Again, by induction, we assume that the $\Xi$-introduction step can
  be applied to the previous steps.

  The axioms of the Hilbert-style system, when rewritten with $U
 \rightarrow V$ for $FUV\equiv\lambda x.\Xi U(\lambda y.V(xy))$
  are:
\[\eqalign{
 \vdash\ldots\  
 &[(A\rightarrow A)\,{\bf I}]\cr
 \vdash\ldots\ 
 &[(A\rightarrow B\rightarrow A)\,{\bf K}]\cr
 \vdash\ldots\ 
 &[((A\rightarrow B
\rightarrow C)\rightarrow (A\rightarrow B)\rightarrow A
\rightarrow C)\,{\bf S}]\cr
  }
\]
  where $\ldots$ represent conditions involving ${\bf L}$ on $A,B$ and
  $C$.

  These are type assignment statements for ${\bf I}$, ${\bf K}$ and
  ${\bf S}$.

  It might be thought that this same technique could be employed for
  PTSs, using type assignment statements for ${\bf I}$, ${\bf K}$ and
  ${\bf S}$, of the form $\vdash(\ldots {\bf I})\:(\ldots A\rightarrow
  A)$ etc and with hypotheses of the form $x\:A$.  This however may
  not work.

  If we have a PTS with $(c:s_1)\in\mathcal A$ and can prove $x\:c
  \vdash B\:s_2$ and/or $x\:c\vdash M\:B$, perhaps with $M\equiv x$,
  $c \equiv B$, it may be that (product) and (abstraction) cannot be
  applied because $(s_1, s_2, s_3)\not\in\mathcal R$ for any $s_3$.

  This does not mean that $x\:c$ can never be cancelled. We may
  obtain:
\[x\:c\vdash P\:(\Pi y{:}D.E)\quad\mbox{and}\quad x\:c\vdash Q\:D\]
  where $x\:c$ cannot be cancelled, as, even if we have
\[x\:c\vdash (\Pi y{:}D.E):s_2\quad\mbox{and}\quad x\:c\vdash D\:s_3,\]
  $(s_1, s_2, s_4)$ and $(s_1, s_3, s_5)$ may not be in $\mathcal R$
  for any $s_4, s_5\in\mathcal S$. However if
\[x\:c\vdash PQ:E [y:=Q]\quad\mbox{,}\quad x\:c\vdash E[y:=Q]:s_6
 \quad\mbox{and also}\quad(s_1, s_6, s_7)\in\mathcal R,
\]
  so that $\vdash(\Pi x{:}c.PQ):s_7$, we can cancel $x\:c$ to give
\[\vdash(\lambda x{:}c.PQ)\:(\Pi x{:}c.E[y:=Q]).\]

  This PTS therefore does have theorems not in ${\mathcal A}$, but it
  is hard to determine the HPTS corresponding to it.

\section{Supersorted PTSs}\label{S:SupSort}

 \noindent PTSs that have equivalent HPTSs are $\lambda^\ast$ and
  $\lambda $Coq (both versions), but these belong to a larger class
  that has the following property{:}

\begin{defi}[Supersorted]\label{D:supers}

  A PTS is said to be \emph{supersorted} if:
\[(\forall c\in {\mathcal C})(\exists s\in {\mathcal S})\ (c:s)\in
  {\mathcal A}
\quad\mbox{and}\quad
  (\forall s_1, s_2\in {\mathcal S})\ (\exists s_3\in {\mathcal
S})\ (s_1,s_2,s_3)\in {\mathcal R}.
\]
\end{defi}

  For supersorted PTSs (abstraction) can be simplified.

\begin{thm}\label{T:sixteen} In every supersorted PTS
  (abstraction) can be replaced by:
\[\infer
{\Gamma\vdash(\lambda x{:}A.M)\:(\Pi x{:}A.B)}{\Gamma, x\:A\vdash
  M\:B}\.
\]
\end{thm}

\proof If
\[\Gamma, x\:A\vdash M\:B\]
  by Lemmas \ref{L:seven}(ii) and \ref{L:five} we have, for some
  $s_1,s_2\in\mathcal S$:
\[\Gamma\vdash A\:s_1\quad\mbox{and either}\quad
 \Gamma, x\:A\vdash B\:s_2
 \quad\mbox{or}\quad B=s\enspace\mbox{for some}\enspace s\in\mathcal S\.
\]
  If the PTS is supersorted we have, for some $s_2$, $(s\:s_2)\in
 \mathcal A$ in the latter ($B=s$) case, and so the result of the
  former case by Lemma \ref{L:seven}(i).

  Hence by (product) and supersortedness we have, for some $s_3\in
  {\mathcal S}$,
\[\Gamma\vdash (\Pi x{:}A.B)\:s_3\]
  and by (abstraction) we have
\[\Gamma\vdash(\lambda x{:}A.M)\:(\Pi x{:}A.B).\]

  For a supersorted PTS $\lambda X$ we define a corresponding HPTS
  $\lambda X^h$, which in Theorem \ref{T:twentyeight} is shown to be
  equivalent to $\lambda X^h$.

\begin{defi}[{Corresponding HPTS}]\label{D:ssHPTS}

  If $\lambda X$ is a supersorted PTS with specification $({\mathcal
    S}, {\mathcal A}, {\mathcal R})$ the \emph{corresponding} HPTS
  $\lambda X^h$ has specification $({\mathcal S}, {\mathcal
    A},{\mathcal B})$, with as members of ${\mathcal B}$ the following
  theorems of $\lambda X$:

\noindent{\bf Axiom \hbox to21 pt{$\bf\Pi 1$\hfil}}%
  $\vdash^h[\lambda u{:}s_1 .\lambda v{:}(\Pi x{:}u
  . s_2)\.\Pi x{:}u.vx]\:[\Pi u{:}s_1.\Pi v{:}(\Pi
  x{:}u. s_2).s_3]$ for $(s_1,s_2,s_3)\in\mathcal R$.\hfill

\noindent{\bf Axiom \hbox to21 pt{$\bf I1$\hfil}}%
  $\vdash^h[\lambda x{:}s_1.\lambda y{:}x.y]:[\Pi x{:}s_1.\Pi y{:}x.x]$.\hfill

\noindent{\bf Axiom \hbox to21 pt{$\bf K1$\hfil}}%
  $\vdash^h[\lambda x{:}s_1 .\lambda
  y{:}s_2.\lambda z{:}x . \lambda u{:}y . z]:[\Pi x{:}s_1 .\Pi
  y{:}s_2 .\Pi z{:}x.\Pi u{:}y.x]$.\hfill

\noindent{\bf Axiom \hbox to21.6 pt{$\bf S1$\hfil}}%
  $\vdash^h[\lambda u{:}s_1.\lambda
  v{:}(\Pi x{:}u.s_2).\lambda t{:}(\Pi x{:}u.(\Pi y{:}vx.s_3)).\lambda
  w{:}(\Pi x{:}u.\Pi y{:}vx.txy)$.\hfill

\noindent\phantom{\bf Axiom \hbox to21 pt{$\bf S1$\hfil}$\vdash^h[$}%
  $\lambda z{:}(\Pi x{:}u.vx).\lambda x{:}u. wx(zx)]\:$\hfill 

\noindent\phantom{\bf Axiom \hbox to21 pt{$\bf S1$\hfil}$\vdash^h{}$}%
  $[\Pi u{:}s_1.\Pi v{:}(\Pi x{:}u.s_2).\Pi t:(\Pi x{:}u .(\Pi y{:}vx.s_3)).\Pi w{:}(\Pi x{:}u.\Pi y{:}vx.txy).$\hfill

\noindent\phantom{\bf Axiom \hbox to21 pt{$\bf S1$\hfil}$\vdash^h[$}%
  $\Pi z{:}(\Pi x{:}u.vx).\Pi x{:}u.tx(zx)]$.\hfill

\noindent and additional axioms of $\mathcal B$ generated by (I) and
  (II):
\begin{enumerate}[(I)]
\item If $ (M\:A)\in\mathcal B$, $A\notin {\mathcal S}$ and $A'$ is
  obtained from $A$ by replacing any second occurrence of an $s_i$ in
  $A$ by any $s_j$ not in $A$, then if $\vdash A'\:s$ for $s\in
  {\mathcal S}$, $ (A'\:s)\in\mathcal B$. Any conditions on $(M\:A)
 \in\mathcal B$ not required in the proof of $\vdash A'\:s$ are not
  part of the new axiom.
\item If $([\lambda x_1{:}A_1\ldots\lambda x_{i-1}{:}A_{i-1}
 \.\Pi x_i{:}A_i.B]\:[\Pi x_1{:}A_1\ldots\Pi x_{i-1}{:}
  A_{i-1}\.s])\in\mathcal B$ and $s'\in\mathcal B$ satisfies
  $\vdash(\lambda x_1{:}A_1\ldots\lambda x_i{:}A_i.B)\:(\Pi x_1{:}A_1
 \ldots\Pi x_i{:}A_i .s')$, then
\[([\lambda x_1{:}A_1\ldots\lambda x_i{:}A_i.B]\:%
  [\Pi x_1{:}A_1\ldots\Pi x_i{:}A_i.s'])\in {\mathcal B}.
\]
\end{enumerate}
\end{defi}

 \noindent{\bf Note.}\ The $s_1,s_2,\dots .$ in Axioms I1, K1 and S1 are
  sort variables that can be replaced by arbitrary elements of
  ${\mathcal S}$. In the axioms generated by (I) and (II) there are
  restrictions on the sorts that can be substituted for such variables
  based on the PTS provability of the judgements mentioned.

  Given a PTS $\lambda$X, we will assume below that $\lambda X^h$ is
  the corresponding HPTS.

\begin{thm}\label{T:seventeen} If, for a supersorted PTS, $\vdash^h
  M\:A$ then $\vdash M\:A$.
\end{thm}

\proof By induction on the derivation of  $\vdash^h M\:A$.

  If $\vdash^h M\:A$ is one of the axioms of $\mathcal A$, $\Pi1$, I1,
  K1 or S1, or is generated by (I) or (II), we have $\vdash M\:A$.

  The (application) and (conversion) cases follow from the induction
  hypothesis.

  The (subject reduction) and (type reduction) cases follow from the
  induction hypothesis and Lemma \ref{L:six}.\qed

\begin{lem}\label{L:eighteen} In a HPTS corresponding to a supersorted PTS,
\begin{enumerate}[\em(i)]
\item If $(M\:A)\in\mathcal A\cup\mathcal B$ then there is an $s
 \in\mathcal S$ such that $(A\:s)\in\mathcal A\cup\mathcal B$.
\item If $([\lambda x_1{:}A_1\ldots\lambda x_{i-1}{:}A_{i-1}
 \.\Pi x_i{:}A_i.B]\:[\Pi x_1{:}A_1\ldots\Pi x_{i-1}{:}
  A_{i-1}\.s])\in\mathcal B$, there is an $s'\in\mathcal S$
  such that $ ([\lambda x_1{:}A_1\ldots\lambda x_i{:}A_i.B]\:[\Pi
    x_1{:}A_1\ldots\Pi x_i{:}A_i .s'])\in\mathcal B$.
\end{enumerate}
\end{lem}

\proof\hfill

\begin{enumerate}[(i)]
\item If $A\in\mathcal S$ this follows by supersortedness.

  \noindent If $(M\:A)\in\mathcal B$ and $A\notin\mathcal S$, we have
  $\vdash M\:A$ by Theorem \ref{T:seventeen} and $\vdash A\:s$, for
  some $s\in\mathcal S$, by Lemma \ref{L:five}.

  \noindent Hence $(A\:s)\in\mathcal B$ by (I).

\item If $([\lambda x_1{:}A_1\ldots\lambda x_{i-1}{:}A_{i-1}\.\Pi
  x_i{:}A_i.B]\:[\Pi x_1{:}A_1\ldots\Pi x_{i-1}{:}A_{i-1}\.s])
 \in\mathcal B$, by Theorem \ref{T:seventeen} and Lemma
  \ref{L:four}(iv) and (iii) we have, for some $s'\in\mathcal S$,
\[x_1\:A_1\ldots x_i\:A_i\vdash B\:s'.\]
  (abstraction) and (II) then give the result.\qed
\end{enumerate}

 \noindent We also need an extension of $\lambda X^h$ that allows hypotheses.

\begin{defi}[$\lambda X^{h+}$]\label{D:lam}
  If $\lambda X$ is a PTS, $\lambda X^{h+}$ has all the postulates of
  $\lambda X^h$, also with nonempty contexts, and the (start) and
  (weakening) rules of $\lambda X$.
\end{defi}

\begin{lem}\label{L:nineteen}
  $\vdash^{h+} M\:A\Leftrightarrow\vdash^h M\:A.$
\end{lem}

\proof Immediate because in a derivation of $\vdash^{h+} M\:A$ no
  (start) or (weakening) rule can be used. No nonempty context can be
  emptied in $\lambda X^{h+}$.\qed

  The extra axioms of ${\mathcal B}$ generated by (I) we will need in
  the proof of the Correctness of Types Lemma for $\lambda X^{h+}$ (if
  $\Gamma\vdash^{h+} M\:A$ then $\Gamma\vdash^{h+} A\:s$ for some $s$).

  Those generated by (II) we need in the proof of (abstraction) to
  show that, if we have $\Gamma\vdash^{h+}(\Pi x{:}C.D):s_3$, we also
  have for $(s_1, s_2, s_3)\in {\mathcal R}$, $\Gamma\vdash^{h+}
  C:s_1$ and $\Gamma\vdash^{h+}(\lambda x{:}C.D)\:(\Pi x{:}C.s_2)$,
  where the derivation of the latter is no longer than the derivation
  of $\Gamma\vdash^{h+}(\Pi x{:}C.D):s_3$. The ``no longer than" is
  needed for proof by induction to work.

  Many of the axioms are, in a sense, superfluous.  We can for
  example, prove axioms $\Pi 4,\Pi 8$ and $\Pi 11$ (below) from Axiom
  K1 and Axiom $\Pi 5$ from Axioms $\Pi 1$ and $\Pi 4$.  However,
  using fewer axioms can mean that the derivation of a
  $\Gamma\vdash^{h+}(\lambda x{:}C.D)\:(\Pi x{:}C.s_2)$ is longer than
  that of $\Gamma\vdash^{h+}(\Pi x{:}C.D{)}:s_3$.

  To illustrate that the axioms, generated by (I) and (II) above, form
  finite sets, we list all the ones generated by Axiom $\Pi 1$($s_i'$
  below is such that $(s_i:s_i' )\in\mathcal A$). There are another six
  I axioms, another sixteen K axioms and twentynine more S axioms.

\noindent{$\bf\Pi 2\phantom{0}$}\quad $\vdash^h [\Pi u{:}s_1  .
 \Pi v{:}(\Pi x{:}u.s_2).s_3]:s_4$.

\noindent{$\bf\Pi 3\phantom{0}$}\quad $\vdash^h [\lambda u{:}s_1 .\Pi
  v{:}(\Pi x{:}u.s_2).s_3]\:[\Pi u{:}s_1.s_4]$.

\noindent{$\bf\Pi 4\phantom{0}$}\quad $\vdash^h [\lambda u{:}s_1.s_2]
\:[\Pi u{:}s_1.s_2']$.

\noindent{$\bf\Pi 5\phantom{0}$}\quad $\vdash^h [\Pi u{:}s_1 . s_2]\:s_3$.

\noindent{$\bf\Pi 6\phantom{0}$}\quad $\vdash^h [\lambda u{:}s_1
  .\lambda v{:}(\Pi x{:}u.s_2) .\lambda x{:}u.vx]\:[\Pi u{:}s_1
  .\Pi v{:}(\Pi x{:}u.s_2) .\Pi x{:}u.s_2]$.

\noindent{$\bf\Pi 7\phantom{0}$}\quad $\vdash^h [\Pi u{:}s_1 .\Pi
  v{:}(\Pi x{:}u.s_2) .\Pi x{:}u . s_3]\:s_4$.

\noindent{$\bf\Pi 8\phantom{0}$}\quad $\vdash^h [\lambda u{:}s_1.
 \lambda v{:}(\Pi x{:}u.s_2).s_3]\:[\Pi u{:}s_1.\Pi v{:}(\Pi
  x{:}u.s_2).s_3']$.

\noindent{$\bf\Pi 9\phantom{0}$}\quad $\vdash^h [\lambda u{:}s_1
  .\Pi v{:}(\Pi x{:}u.s_2).\Pi x{:}u.s_3]\:[\Pi
  u{:}s_1.s_4]$.

\noindent{$\bf\Pi 10$}\quad $\vdash^h [\lambda u{:}s_1  .\lambda
v{:}(\Pi x{:}u.s_2) .\Pi x{:}u.s_3]\:[\Pi u{:}s_1 . \Pi v{:}(\Pi
x{:}u.s_2) . s_4]$.

\noindent{$\bf\Pi 11$}\quad $\vdash^h [\lambda u{:}s_1  .\lambda
v{:}(\Pi x{}u.s_2)  .\lambda x{:}u.s_3]\:[\Pi u{:}s_1.\Pi v{:}(\Pi
x{:}u.s_2).\Pi x{:}u. s_3']$.

  The axiom required by (I) for $\Pi1,\Pi8$ and $\Pi10$ is $\Pi2$,
  for $\Pi3,\Pi4$ and $\Pi9,\Pi5$, for $\Pi11$, $\Pi7$ and for
  $\Pi6$ the instance of $\Pi 7$ where $s_2 = s_3$. The axiom required
  by (II) for $\Pi1$ is $\Pi6$, for $\Pi2$ is $\Pi3$, for $\Pi3$ is
  $\Pi8$, for $\Pi5$ is $\Pi4$, for $\Pi7$ is $\Pi9$, for $\Pi9$ is
  $\Pi10$ and for $\Pi10$ is $\Pi11$.

  Each axiom is an axiom scheme in the sense that it is an axiom for
  all $s_1,s_2,\dots $ for which it is provable in $\lambda X$. Thus most
  axioms (not $\Pi$6) have some restrictions, other than $s_i:s'_i\in
 \mathcal A$, for example $(s_1,s_2',s_5), (s_5,s_3',s_4)\in
 \mathcal R$ in $\Pi 3$. Some of these restrictions will appear in
  (the proofs of) some of the lemmas for $\lambda X^{h+}$ below.

  We will show later that in $\lambda X^{h +}$, for a suitable
  $\lambda X$, (product) and (abstraction) are admissible and that the
  theorems of $\lambda X^{h +}$ are exactly those of $\lambda X^h$ and
  $\lambda X$.

\section{The Correctness of Types Lemmas for \texorpdfstring{$\lambda
    X^{h +}$}{lambda X{h+}}}\label{S:CTL}

 \noindent To state and prove some preliminary lemmas we need some
  definitions.

\begin{defi}[major premise chain]\label{D:mpc}
  A \emph{major premise chain (mpc)} in a derivation is a sequence of
  judgements starting with one formed by a (start) rule or an
  axiom. The remaining judgements of the chain are obtained by
  (weakening), (application) or (conversion), with the previous
  judgement as major premise, or by (subject reduction) or (type
  reduction). The minor premises in (weakening), (application) and
  (conversion) rules for which the major premises are in an mpc, will
  be called the \emph{minor premises attached to the mpc}.
\end{defi}

  The final judgement of an mpc that is not a proper part of a larger
  mpc, must be the final judgement in a derivation, a judgement that
  is the premise for a (start) rule or the minor premise in a
  (weakening), (application) or (conversion) rule.

  Any derivation is therefore made up of linked mpcs.

\begin{defi}\label{D:losh}
   An mpc is said to be \emph{long} if it starts with an axiom of the form
\begin{equation}\label{EQ:eighteen}
 \vdash^{h+}[\lambda x_1{:}A_1\dots\lambda x_n{:}A_n.N]\:
  [\Pi x_1{:}A_1\dots\Pi x_n{:}A_n.A_{n+1}]
\end{equation}
  where $n> 0$, $N$ is one of $x_1,\dots ,x_n$ or is formed by
  application from (some of) $x_1,\dots ,x_n$ and the mpc has at least
  $n$ (application) steps and (subject reduction) steps that reduce
  all of the $n$ $\lambda x_i$ redexes. An mpc is \emph{short}
  otherwise.

  A derivation is \emph{short} if it has no long mpcs and \emph{long}
  otherwise.
\end{defi}

\begin{defi}[Application Length - alength]\label{D:alength}
  The \emph{application length} or \emph{alength} of a derivation is
  its number of (application) steps, where steps in identical minor
  premises in the derivation, are counted only once.

  A derivation of lesser alength than another will be called
  \emph{ashorter}, one of greater alength \emph{alonger}.
\end{defi}

 \noindent {\bf Note.}\ One derivation of a judgement may be shorter
  (in length) than another without being short.

\begin{lem}\label{L:twenty} If the final mpc in a derivation of
\begin{equation}\label{EQ:nineteen}
 \Gamma\vdash^{h+} M\:A.
\end{equation}
  is long, it starts with an axiom of the form
  \emph{(\ref{EQ:eighteen})} and the (subject reduction) step that
  reduces the $\lambda x_n$ redex comes directly after the $n$th
  (application) step, then that derivation of
  \emph{(\ref{EQ:nineteen})} can be replaced by an ashorter one.
\end{lem}

\proof This has to be proved for each of the axioms of $\mathcal B$
  that is of this form. We will prove it for Axiom {S8}, below, the
  proofs for other axioms are similar.
\bigskip

\noindent{\bf{S8}}\quad 
$[\lambda y_1{:}s_q.%
  \lambda y_2{:}(\Pi x{:}y_1.s_r).%
  \lambda y_3{:}(\Pi x{:}y_1.\Pi y{:}y_2x.s_t).%
  \lambda y_4{:}(\Pi x{:}y_1.\Pi y{:}y_2x.y_3xy).$

\noindent\phantom{\bf{S8}[}\quad
$\lambda y_5{:}(\Pi x{:}y_1.y_2x).
 \lambda y_6{:}y_1.y_3y_6(y_5y_6)]\:$\hfill

\noindent\phantom{\bf{S8}}\quad
$[\Pi y_1{:}s_q.\Pi y_2{:}(\Pi x{:}y_1.s_r).\Pi y_3{:}(\Pi x{:}y_1.\Pi y{:}y_2x.s_t).\Pi y_4{:}(\Pi x{:}y_1.\Pi y{:}y_2x.y_3xy).$

\noindent\phantom{\bf{S8}[}\quad
$\Pi y_5{:}(\Pi x{:}y_1.y_2x).%
 \Pi y_6{:}y_1. s_t)]$.
\bigskip

 \noindent Let the minor premises in the six (application) steps
 involving $\Pi y_1$ to $\Pi y_6$ in the long derivation of
 (\ref{EQ:nineteen}) be, for $1\le i\le 6$:
\[\Gamma_i\vdash^{h+} Y_i:E_i\]
  where $E_1=_{\beta}s_q$, $E_2=_{\beta}\Pi x{:}Y_1.s_r$, $E_3
  =_{\beta}\Pi x{:}Y_1.\Pi y{:}Y_2x.s_t$, $E_4=_{\beta} (\Pi
  x{:}Y_1.\Pi y{:}Y_2x.Y_3xy)$, $ E_5 =_{\beta}\Pi x{:}Y_1.Y_2x$,
  $E_6=_{\beta} Y_1, A=_{\beta}s_t$ and $Y_3Y_6(Y_5Y_6)\to\hskip -7pt
 \to_{\beta} M$.

  Then for some $Y_1^*$, $Y_1\to\hskip -7pt\to_{\beta} Y_1^*$ and
  $E_6\to\hskip -7pt\to_{\beta} Y_1^*$ and for some $R[x]$, $Y_2x
 \to\hskip -7pt\to_{\beta} R[x]$, $E_3\to\hskip -7pt\to_{\beta}
 \Pi x{:}Y_1^*.\Pi y{:}R[x].s_3$ and $E_5\to\hskip -7pt\to_{\beta}
 \Pi x{:}Y_1^*. R[x]$.

  Note that as contexts can only grow, each $\Gamma_i$ for $1\le i <
  6$ is an initial segment of $\Gamma_{i+1}$

  Now by (weakening), (subject reduction), (type reduction) and just
  three (application) steps we get from three of these minor premises:
\[\eqalign{
 \Gamma_6 &\vdash^{h+} Y_3Y_6\:\Pi y{:}R[X_i].s_t,\cr
 \Gamma_6 &\vdash^{h+} Y_5Y_6\:R[X_i]\cr
 \!\hbox to170 pt{and so\hfill}&\hbox to262pt{\hfill}\cr
 \Gamma_6 &\vdash^{h+} Y_3Y_6(Y_5Y_6):s_t\cr
  }
\]
  which, as $Y_3Y_6(Y_5Y_6)\to\hskip -7pt\to_{\beta} M$, gives
  (\ref{EQ:nineteen}).

  We now have a new derivation of (\ref{EQ:nineteen}), which, given
  that any (application)s in the two uses of $\Gamma_6\vdash Y_6\:E_6$
  are counted only once, has fewer (application)s, and so is ashorter
  than, the old derivation of (\ref{EQ:nineteen}).

\begin{lem}[Shortness Lemma for HPTS$^+$]\label{L:twentyone}
  Every valid judgement in a HPTS$^+$ has a short derivation.
\end{lem}

\proof We prove this by showing that for every long
  derivation there is an ashorter derivation of the same judgement.

  Assume that the following is the part of a long mpc, in a long
  derivation, up to the $\lambda x_n$ reduction, together with the
  minor premises used in the n (application) steps.

{\small
$$\vdash^{h+} T_1\:\Pi x_1{:}B_0.C_0\hspace{1cm}$$\vspace{-3mm}
$$\vdots\hspace {2cm}$$
$$\Gamma_1\vdash^{h+} T_1\:\Pi x_1{:}B_1.C_1\hspace {2cm}\Gamma_1\vdash^{h+} X_1:B_1 \vspace{-2mm}$$
\vspace{-2mm}\hspace{3cm}\hbox{\vrule height .4pt width 10cm}
$$\Gamma_1\vdash^{h+} T_1X_1\:C_1[x_1:=X_1]$$
\hspace {6.2cm}\vdots
$$\Gamma_2\vdash^{h+} T_2\:\Pi x_2{:}B_2.C_2\hspace {2cm}\Gamma_2\vdash^{h+} X_2:B_2\vspace{-2mm}$$
\vspace{-2mm}\hspace{3cm}\hbox{\vrule height .4pt width 10cm}
$$\Gamma_2\vdash^{h+} T_2X_2\:C_2[x_2:=X_2]$$
\hspace {6.2cm}\vdots
$$\Gamma_n\vdash^{h+} T_n\:\Pi x_n{:}B_n.C_n\hspace {2cm}\Gamma_n\vdash^{h+} X_n\:B_n\vspace{-2mm}$$
\vspace{-2mm}\hspace{3cm}\hbox{\vrule height .4pt width 10cm}
$$\Gamma_n\vdash^{h+} T_nX_n\:C_n[x_n:=X_n] $$
\vspace{-2mm}\hspace{3cm}
$$\vdots\hspace {1cm}$$
$$\Gamma_{n+1}\vdash^{h+} (\lambda x_n{:}A_n'.N')X_n'Y_1\dots Y_k\:D\vspace{-2mm}$$
\vspace{-2mm}\hspace{3cm}\hbox{\vrule height .4pt width 10cm}
$$\Gamma_{n+1}\vdash^{h+} N'[x_n:=X_n']Y_1\dots Y_k\:D\vspace{-2mm}$$
}

 \noindent Here $\vdash^{h+} T_1 :\Pi x_1{:}B_0.C_0$ is an axiom of
  the form (\ref{EQ:eighteen}) with $N$ made up of (some of)
  $x_1,\dots ,x_n$, $ T_n\to\hskip -7pt\to_{\beta}\lambda
  x_n{:}A_n'.N'$ and $X_n\to\hskip -7pt\to_{\beta} X_n'.$ The
  first, second and $n$th of the $n$ or more (application)s and the
  (subject reduction) contracting the $\lambda x_n$ redex are
  explicitly shown. The steps after the $n$th (application) only alter
  $T_n$ by reducing it, so steps can be permuted so that the $\lambda
  x_n$ reduction takes place straight after the $n$th (application)
  step as follows:
\medskip

{\small
\hspace {3.6cm}$\Gamma_n\vdash^{h+} T_n\:\Pi x_n{:}B_n.C_n\vspace{-5mm}$

\vspace{2mm}\hspace{3.6cm}\hbox{\vrule height .4pt width 5.2cm}
\vspace{-2mm}
$$\qquad\quad\quad\Gamma_n\vdash^{h+} \lambda x_n{:}A_n'.N'\:\Pi 
x_n{:}B_n.C_n\hspace {2cm}\Gamma_n\vdash^{h+} X_n\:B_n\vspace{-2mm}$$
\vspace{-2mm}\hspace{3cm}\hbox{\vrule height .4pt width 10cm}
$$\Gamma_n\vdash^{h+} (\lambda x_n{:}A_n'.N')X_n\:C_n[x_n:=X_n]\vspace{-2mm}$$
\vspace{-2mm}\hspace{4cm}\hbox{\vrule height .4pt width 9cm}
$$\Gamma_n\vdash^{h+} N'[x_n:=X_n]\:C_n[x_n:=X_n]\vspace{-2mm}$$
\hspace {7cm}\vdots
$$\Gamma_{n+1}\vdash^{h+} N'[x_n:=X_n']Y_1\dots Y_k\:D\vspace{-2mm}$$
}

 \noindent This new derivation is no alonger than the original, but
  the part up to $\Gamma_n\vdash^{h+} N'[x_n:=X_n']\:C_n[x_n:=X_n]$
  is long and can be replaced, by Lemma \ref{L:twenty}, by an ashorter
  derivation, so the whole derivation becomes ashorter. (If the
  derivation had identical mpcs to the above, which were all minor
  premises in the same mpc, all would have to be altered as above to
  ensure that the new derivation is not alonger than the old.)

  In the remaining lemmas and theorems we use a different measure of
  length of a derivation, where ``similar" subderivations are counted
  only once.

\begin{defi}[Similar]\label{D:sim}
  Two derivations are said to be  \emph{similar} if they are identical
  or one, in its final mpc, starts with an axiom of $\mathcal B$ of
  the form (\ref{EQ:eighteen}), and the other differs only in that its
  final mpc starts with an axiom of $\mathcal B$ generated from the
  other by one or more applications of (II).
\end{defi}

  We now define:

\begin{defi}[Similarity Length - slength]\label{D:simlen}
  The  \emph{similarity length} (or  \emph{slength}) of a derivation is
  given by:
\begin{enumerate}[(i)]
\item the number of (application) steps,
\item the number of (conversion), (start) and (weakening) steps.
\end{enumerate}
  Similar derivations ending in the two premises of a (weakening)
  step, are counted only once.

  A derivation of lesser slength than another will be said to be
  \emph{sshorter} and one of greater slength as  \emph{slonger}.
\end{defi}

\begin{lem}\label{L:twentytwo} Given, for $s'\in {\mathcal S}$, a
  short derivation of:
\begin{equation}\label{EQ:twenty}
 \Gamma\vdash^{h+}\Pi x{:}B.C:s',
\end{equation}
  there is, for some $s\in\mathcal S$, a derivation, no longer or
  slonger than that of \emph{(\ref{EQ:twenty})}, of
\begin{equation}\label{EQ:twentyone}
 \Gamma\vdash^{h+}\lambda x{:}B.C:\Pi x{:}B.s.
\end{equation}
\end{lem}

\proof Consider the first judgement in the final mpc in a short
  derivation of (\ref{EQ:twenty}). This cannot be an axiom of ${\mathcal
  A}$ or be formed by a (start) rule, so it is an axiom of ${\mathcal
  B}$ of the form (\ref{EQ:eighteen}), where $N\equiv\Pi x{:}B'.C'$
  and $A_{n+1}\equiv s'$.

  If in this mpc we replace this axiom by the one generated from it by
  (II), then using exactly the same steps and minor premises we obtain
  a derivation of (\ref{EQ:twentyone}) of the same length.

  In this final mpc there are no (weakening) steps in which the
  premises are similar, until perhaps after the last (application)
  step, as, until then, no type can be in ${\mathcal S}$. If, at that
  stage, (\ref{EQ:twenty}) is formed by one or more (weakening) steps
  (and perhaps (subject reduction)) from $\Gamma^-\vdash^{h+}\Pi
  x{:}B^0.C^0:s'$ and similar minor premises such as $\Gamma^-
 \vdash^{h+} D:s'$, these are counted only once each in the
  slength. In the derivation obtained by changing the axiom, the above
  derivations remain similar and so the slength of the derivation
  remains the same.\qed

\begin{lem}[Correctness of Types for HPTS$^+$]\label{L:twentythree}
  If $\lambda$X is supersorted and
\begin{equation}\label{EQ:twentytwo}
 \Gamma\vdash^{h+} M\:A,
\end{equation}
  then, for some $s\in\mathcal S$, $A\equiv s$ or there is a short
  derivation, of slength no more than that of a short derivation of
  \emph{(\ref{EQ:twentytwo})}, of
\begin{equation}\label{EQ:twentythree}
 \Gamma\vdash^{h+}A\:s.
\end{equation}
\end{lem}

\proof By induction on the number k, of judgements in the final mpc of
  a short derivation of $\Gamma\vdash^{h+} M\:A$, where $A\not\in
 \mathcal S$.

If k=1 and (\ref{EQ:twentytwo}) comes by a (start) rule from
\begin{equation}\label{EQ:twentyfour}
 \Gamma^-\vdash^{h+} A\:s,
\end{equation}
  where $M\equiv x$ and $\Gamma\equiv\Gamma^-, x\:A$,
  (\ref{EQ:twentythree}) comes from two copies of
  (\ref{EQ:twentyfour}) and (weakening). The two derivations of
  (\ref{EQ:twentyfour}) are counted only once, so this derivation of
  (\ref{EQ:twentythree}) is no slonger than that of
  (\ref{EQ:twentytwo}).

  If (\ref{EQ:twentytwo}) is an axiom we have (\ref{EQ:twentythree})
  by (I) or by supersortedness.

  We now assume k $\ge$ 2.

  If (\ref{EQ:twentytwo}) comes from $\Gamma'\vdash^{h+} M\:A$ and
  $\Gamma'\vdash^{h+} B\:s'$, by (weakening), where $s'\in {\mathcal
    S}$, $A\not\in {\mathcal S}$ and $\Gamma =\Gamma', x\:B$, these
  derivations are both counted in the slength of the derivation of
  (\ref{EQ:twentytwo}). We have, by the induction hypothesis, $\Gamma'
 \vdash^{h+} A\:s$, by a derivation no slonger than that of $\Gamma'
 \vdash^{h+} M\:A$, for some $s\in {\mathcal S}$ and we obtain
  (\ref{EQ:twentythree}) by (weakening), by a derivation that is no
  slonger than that of (\ref{EQ:twentytwo}).

  If (\ref{EQ:twentytwo}) comes from $\Gamma\vdash^{h+} N\:A$, by
  (subject reduction), we have (\ref{EQ:twentythree}) by a derivation
  no slonger than that of (\ref{EQ:twentytwo}).

  If (\ref{EQ:twentytwo}) comes from $\Gamma\vdash^{h+} M\:B$, by
  (type reduction), we have $\Gamma\vdash^{h+} B\:s$ by the induction
  hypothesis and (\ref{EQ:twentythree}) by (subject reduction), by a
  derivation no slonger than that of (\ref{EQ:twentytwo}).

  If (\ref{EQ:twentytwo}) comes from $\Gamma\vdash^{h+} M\:B$, by
  (conversion), we have (\ref{EQ:twentythree}) by a derivation
  sshorter than that of (\ref{EQ:twentytwo}).

  If (\ref{EQ:twentytwo}) comes from $\Gamma\vdash^{h+} P:\Pi
  x{:}B.C$ and $\Gamma\vdash^{h+} Q:B$, where $M\equiv PQ$ and
  $A\equiv C[x:=Q]$, by (application), we have by the induction
  hypothesis, $\Gamma\vdash^{h+}\Pi x{:}B.C:s'$ for some $s'\in
 \mathcal S$, by a derivation no slonger than that of $\Gamma
 \vdash^{h+} P:\Pi x{:}B.C$. Then by Lemma \ref{L:twentytwo} we have
  $\Gamma\vdash^{h+}\lambda x{:}B.C :\Pi x{:}B.s$, for some $s\in
  {\mathcal S}$ by a derivation no slonger than that of $\Gamma
 \vdash^{h+} P:\Pi x{:}B.C$. Then by (application) using $\Gamma
 \vdash^{h+} Q:B$ we have (\ref{EQ:twentythree}) by a derivation no
  slonger than that of (\ref{EQ:twentytwo}).\qed

\begin{lem}[Start Lemma for HPTS$^+$]\label{L:twentyfour}
  If 
\begin{equation}\label{EQ:twentyfive}
 \Gamma, x\:A\vdash^{h +} M\:B
\end{equation}
  then, for some $s\in\mathcal S$,
\[\Gamma\vdash^{h +} A\:s.\]
\end{lem}

\proof By an easy induction on the derivation of (\ref{EQ:twentyfive}).\qed

\section{The Equivalence Results}\label{S:ER}

\begin{lem}\label{L:twentyfive}
  If $\lambda X$ is supersorted, (abstraction) is
  admissible in $\lambda X^{h +}.$
\end{lem}

\proof If $\lambda X$ is supersorted we prove that if
\begin{equation}\label{EQ:twentysix}
 \Gamma, x\:A\vdash^{h +} M\:B
\end{equation}
  then
\begin{equation}\label{EQ:twentyseven}
 \Gamma\vdash^{h +}(\lambda x{:}A.M)\:(\Pi x{:}A.B)
\end{equation}
  by induction on the slength of a short derivation of (\ref{EQ:twentysix}).

\noindent{\bf Case 1.}\ (\ref{EQ:twentysix}) comes by (start) (and (type reduction)) from
\[\Gamma\vdash^{h +}A\:s\]
  where $M\equiv x$ and $A\to\hskip -7pt \to_{\beta} B$.

  By Axiom I1 and (application) (and (type reduction)) we have (\ref{EQ:twentyseven}).

\noindent{\bf Case 2.} (\ref{EQ:twentysix}) comes by (weakening) (and reduction) from
\[\Gamma\vdash^{h +}M'\:B'\quad\mbox{and}\quad\Gamma\vdash^{h +}A\:s_2\]
  then by the Correctness of Types Lemma (Lemma \ref{L:twentythree}) or
  supersortedness, for some $s_1$.
\[\Gamma\vdash^{h +} B'\:s_1\]
  and (\ref{EQ:twentyseven}) follows after three (applications) applied to Axiom K1 (and
  reduction).

\noindent{\bf Case 3.}\ (\ref{EQ:twentysix}) comes by (conversion) (and reduction) from
\[\Gamma, x\:A\vdash^{h+} M:C\quad\mbox{,}\quad
 \Gamma, x\:A\vdash^{h+} B'\:s_2\quad\mbox{and}\quad
  C=_{\beta} B'\to\hskip -7pt \to_{\beta}B.
\]
  By the induction hypothesis and (subject reduction) we have:
\begin{equation}\label{EQ:twentyeight}
 \Gamma\vdash^{h +} (\lambda x{:}A.M)\:(\Pi x{:}A.C)
\end{equation}
  and
\[\Gamma\vdash^{h +} (\lambda x{:}A.B)\:(\Pi x{:}A.s_2)\]
  where $\Pi x{:}A.C =_{\beta}\Pi x{:}A.B$.

  We have by Lemma \ref{L:twentythree} applied to (\ref{EQ:twentysix}), for some $s_1$,
\[\Gamma\vdash^{h+} A\:s_1\]
  and, by supersortedness $(s_1,s_2,s_3)\in {\mathcal R}$ for some
  $s_3$, so by Axiom $\Pi 1$ and (subject reduction),
\[\Gamma\vdash^{h +}(\Pi x{:}A.B):s_3,\]
  hence by (\ref{EQ:twentyeight}) and (conversion) we have
  (\ref{EQ:twentyseven}).

\noindent{\bf Case 4}\ (\ref{EQ:twentysix}) comes by (application) (and reduction) from
\begin{equation}\label{EQ:twentynine}
 \Gamma, x\:A\vdash^{h +} P:(\Pi y{:}C.D)
\end{equation}
  and
\begin{equation}\label{EQ:thirty}
 \Gamma, x\:A\vdash^{h +}Q:C
\end{equation}
  where $ PQ\to\hskip -7pt\to_{\beta}M$ and $D[y:=Q]\to\hskip -7pt
 \to_{\beta}B$.

  By the Correctness of Types lemma we have for some $s_4\in{\mathcal
  S}$, by a derivation no slonger than that of (\ref{EQ:twentynine}):
\begin{equation}\label{EQ:thiryone}
 \Gamma, x\:A\vdash^{h+}(\Pi y{:}C.D):s_4
\end{equation}
  now by Lemma \ref{L:twentytwo} we have for some $s_3\in{\mathcal S}$, by a
  derivation no slonger than that of (\ref{EQ:twentynine}), and so sshorter than that
  of (\ref{EQ:twentynine}):
\begin{equation}\label{EQ:thirtytwo}
 \Gamma, x\:A\vdash^{h+}(\lambda y{:}C.D)\:(\Pi y{:}C.s_3)
\end{equation}
  Now by the induction hypothesis applied to (\ref{EQ:twentynine}),
  (\ref{EQ:thirty}) and (\ref{EQ:thirtytwo}) we have:
\begin{equation}\label{EQ:thirtythree}
 \Gamma\vdash^{h+}(\lambda x{:}A.P)\:(\Pi x{:}A.\Pi y{:}C.D)
\end{equation}
\begin{equation}\label{EQ:thirtyfour}
 \Gamma\vdash^{h+}(\lambda x{:}A.Q)\:(\Pi x{:}A.C)
\end{equation}
\begin{equation}\label{EQ:thirtyfive}
 \Gamma\vdash^{h+}(\lambda x{:}A.\lambda y{:}C.D)\:(\Pi x{:}A.\Pi y{:}C.s_3)
\end{equation}
  also by the Correctness of Types Lemma applied to
  (\ref{EQ:thirtyfour}) we have for some $s_4\in{\mathcal S}$
\[\Gamma\vdash^{h+}(\Pi x{:}A.C):s_4\]
  and by Lemma \ref{L:twentytwo} for some $s_2\in {\mathcal S}$
\begin{equation}\label{EQ:thirtysix}
 \Gamma\vdash^{h+}(\lambda x{:}A.C)\:(\Pi x{:}A.s_2)
\end{equation}
  now by Axiom S1, $\Gamma\vdash^{h+}A\:s_1$, (obtained as in Case 3)
  (\ref{EQ:thirtysix}), (\ref{EQ:thirtyfive}), (\ref{EQ:thirtythree}),
  (\ref{EQ:thirtyfour}) and five (application)s, (subject reduction)
  and (type reduction) give (\ref{EQ:twentyseven}). (Note that in
  Axiom S1 $s_1, s_2$ and $s_3$ (here $s_1, s_5$ and $s_2$) can be
  arbitrarily chosen in a supersorted PTS).\qed

\begin{lem}\label{L:twentysix}
  If $\lambda X$ is supersorted (product) is admissible in $\lambda
  X^{h +}$.
\end{lem}

\proof If $\Gamma\vdash^{h+} A\:s_1$, $\Gamma ,x\:A\vdash^{h+} B\:s_2$
 and $(s_1,s_2,s_3)\in {\mathcal R}$, by Lemma \ref{L:twentyfive},
\[\Gamma\vdash^{h+}\lambda x{:}A.B\:\Pi x{:}A.s_2\]
  so by Axiom II1, (application) and (subject reduction) we have
\[\Gamma\vdash\Pi x{:}A.B\:s_3.\eqno{\qEd}\]

  Lemmas \ref{L:twentyfive} and \ref{L:twentysix} show that a theorem that can be proved in $\lambda
  X^{h+}$, using hypotheses, (abstraction) and (product), can also be
  proved in $\lambda X^h$. So:

\begin{thm}\label{T:twentyseven}
  If $\lambda X$ is supersorted it is equivalent to $\lambda X^{h+}$
  in that they have the same valid judgements.
\end{thm}

\proof By Theorem \ref{T:seventeen}, $\lambda X^h$ is a subsystem of
  $\lambda X$. The additional rules of $\lambda X^{h+}$ are rules of
  $\lambda X$, so $\lambda X^{h+}$ is a subsystem of $\lambda X$. The
  extra rules of $\lambda X$ have been shown to be admissible in
  $\lambda X^{h+}$ in Lemmas \ref{L:twentyfive} and \ref{L:twentysix},
  so $\lambda X^{h+}$ and $\lambda X$ have the same valid
  judgements.\qed

\begin{thm}\label{T:twentyeight}
  If $\lambda X$ is supersorted $\lambda X$ and $\lambda X^h$ are
  equivalent in that they have the same theorems.
\end{thm}

\proof It follows from Theorem \ref{T:twentyseven} that $\lambda X$
  and $\lambda X^{h+}$ have the same valid judgements with empty
  contexts i.e. theorems and so by Lemma \ref{L:nineteen} that $\lambda
  X$ and $\lambda X^h$ have the same theorems.\qed

\section {Axioms I, K, S and \texorpdfstring{$\Pi$}{Pi} as Types}\label{S:IKSPi}

 \noindent Axioms I1, K1 and S1 can be rewritten in terms of type
  free combinators (allowing $\eta$-reduction) as:
\medskip

\noindent{\bf Axiom \hbox to21 pt{$\bf I 1$\hfil}}%
  $\vdash{\bf KI}\:(\Pi x{:}s_1 .\Pi y{:}x.x)$.
\smallskip

\noindent{\bf Axiom \hbox to21 pt{$\bf K 1$\hfil}}%
  $\vdash {\bf K(KK)}\:(\Pi x{:}s_1.\Pi y{:}s_2.\Pi z:y .\Pi u{:}x.y)$.
\smallskip

\noindent{\bf Axiom \hbox to21 pt{$\bf S 1$\hfil}}%
  $\vdash {\bf K(K(KS))}\:[\Pi u{:}s_1.\Pi v{:}(\Pi x{:}u.s_2) .\Pi
  t{:}(\Pi x.u.\Pi y{:}vx.s_3)$.

\noindent\phantom{\bf Axiom \hbox to21 pt{$\bf S 1$\hfil}%
  $\vdash{\bf K(K(KS))}\:[$}%
  $\Pi w{:}(\Pi x{:}u.\Pi y{:}vx. txy) .\Pi z{:}(\Pi x{:}u.vx).\Pi
  x{:}u. tx(zx)]$.
\medskip

 \noindent These give the standard types of the combinators (writing
  $A\rightarrow B$ for $\Pi x{:}A.B$ when $x\notin FV(B)$):
{\small\[\eqalign{
 \vdash A\:s_1
&\RIghtarrow\vdash {\bf I}:A\rightarrow A\cr
 \vdash A\:s_1,\vdash B\:s_2 
&\RIghtarrow\vdash {\bf K}:B\rightarrow A\rightarrow B\cr
 \vdash A\:s_1,\vdash B:A\rightarrow s_2,\ 
 \vdash C\:(\Pi x{:}A .\Pi y{:}Bx.s_3) 
&\RIghtarrow\vdash{\bf S}:\Pi w{:}(\Pi
  x{:}A .\Pi y{:}Bx .Cxy) .\cr
&\phantom{\RIghtarrow\vdash{\bf S}:{}}%
 \,\,\Pi z{:}(\Pi x{:}A.Bx) .\Pi x{:}A .Cx(zx)
  }
\]}
  or, as a special case
\small{\[\vdash A\:s_1,\vdash B'\:s_2,\vdash
C'\:A\rightarrow B'\rightarrow s_3\RIghtarrow\vdash{\bf
S}\:(A\rightarrow B'\rightarrow C')\rightarrow (A\rightarrow
B')\rightarrow (A\rightarrow C')
\]}
  If $\Pi x{:}U\.Vx$, where $x\notin FV(UV)$, were represented as
  ${\bf G}UV$ (as it is in $ICL$), Axiom $\Pi 1$ represents the type
  for ${\bf G}$ (or $\lambda uv\.\Pi x{:}u\.vx$):
\[\vdash {\bf G}\:(\Pi u{:}s_1\.\Pi v{:}(\Pi x{:}u\.s_2)\.\
s_3)\quad\mbox{or}\quad\vdash {\bf G}\:(\Pi u{:}s_1\.(u\rightarrow
s_2)\rightarrow s_3).
\]

\section{Identifying \texorpdfstring{$\lambda$}{lambda} and 
\texorpdfstring{$\Pi$}{Pi}}\label{S:LamPi}

  \noindent In the de Bruijn AUTOMATH systems $\Pi$ and $\lambda$ are
  usually identified.  Kameraddine has studied the effect of allowing
  $\beta$-reductions in the (former) $\Pi$ terms in [5].  Doing this
  Axiom {I1} becomes:
\[\vdash {\bf KI}\:(\lambda x{:}s_1\.\lambda y{:}x\.x)\]
  and similarly for the other axioms.  If we write the type in terms
  of combinators we can get (depending on the algorithm)
\[\vdash{\bf KI}\:{\bf S}({\bf KK}){\bf I}\quad\mbox{or}\quad
 \vdash {\bf KI}\:{\bf K}.
\]

\section{Conclusion}\label{S:Conc}

  \noindent We have shown that PTSs come in at least three categories:
  those satisfying ($\$s$) and ($\$\$s$) that have no equivalent HPTS,
  those satisfying ($\$s_1\dots s_n$) that have only a trivial
  equivalent HPTS and supersorted PTSs, such as $\lambda\rbox{Coq}$,
  that have a nontrivial equivalent HPTS. The standard PTSs from the
  literature that we considered all fit into these categories.

\end{document}